\documentstyle[twoside,fleqn,npb,epsfig]{article}

\newcommand{\beq}{\begin{equation}}
\newcommand{\eeq}{\end{equation}}
\newcommand{\barr}{\begin{eqnarray}}
\newcommand{\earr}{\end{eqnarray}}
\newcommand{\ba}{\begin{array}}
\newcommand{\ea}{\end{array}}
\newcommand{\bfp}{\mbox{\boldmath $p$}}
\newcommand{\bfk}{\mbox{\boldmath $k$}}

\newcommand{\la}{\lambda}

\newcommand{\pup}{p^\uparrow}
\newcommand{\aup}{a^\uparrow}
\newcommand{\bup}{b^\uparrow}
\newcommand{\cu}{c^\uparrow}
\newcommand{\qup}{q^\uparrow}
\newcommand{\pdown}{p^\downarrow}

\newcommand{\bdown}{b^\downarrow}
\newcommand{\cd}{c^\downarrow}
\newcommand{\qdown}{q^\downarrow}

\newcommand{\AmS}{{\protect\the\textfont2
  A\kern-.1667em\lower.5ex\hbox{M}\kern-.125emS}}

\title{Single transverse spin asymmetries in 
       inclusive hadron production}

\author{M. Anselmino\address{ Dipartimento di Fisica Teorica, Universit\`a 
        di Torino and \\
        INFN, Sezione di Torino, Via P. Giuria 1, 10125 Torino, Italy},
        M. Boglione\address{ Dept. of Physics and Astronomy, Vrije 
        Universiteit Amsterdam, \\
        De Boelelaan 1081, 1081 HV Amsterdam, The Netherlands}
        and
        F. Murgia\address{ Istituto Nazionale di Fisica Nucleare, 
        Sezione di Cagliari and \\
        Dipartimento di Fisica, Universit\`a di Cagliari,
        C.P. 170, I-09042 Monserrato (CA), Italy}}
            
\begin{document}

\begin{abstract}
A consistent phenomenological approach to the computation of transverse 
single spin asymmetries in inclusive hadron production is presented, based 
on the assumed generalization of the QCD factorization theorem to the case 
in which quark intrinsic motion is taken into account.   
New $\bfk_\perp$ and spin dependent quark distribution and fragmentation 
functions are considered: some of them are fixed by fitting data on
$\pup p \to \pi X$ and predictions are given for single spin asymmetries
in $\ell \pup \to \pi X $ and $\gamma^* \pup \to \pi X$ processes. 
\end{abstract}

\maketitle

\section{General Formalism}
It is well known that perturbative QCD and the factorization theorem 
at leading twist \cite{col} can be used to describe the large $p_T$ 
production of a hadron $C$ resulting from the interaction of two 
polarized hadrons $A$ and $B$
$$
\frac{E_C \, d^3\sigma^{A,S_A + B,S_B \to C + X}} {d^{3} \bfp_C} =
\sum_{a,b,c,d;\{\lambda\}}
\int \frac {dx_a \, dx_b} {16 \pi^2 z \hat s^2}  
$$
\beq
\rho_{\la^{\,}_a, \la^{\prime}_a}^{a/A,S_A} \, f_{a/A}(x_a) \,
\rho_{\la^{\,}_b, \la^{\prime}_b}^{b/B,S_B} \, f_{b/B}(x_b) \label{dsabpol}
\eeq
$$
\hat M_{\la^{\,}_c, \la^{\,}_d; \la^{\,}_a, \la^{\,}_b} \,
\hat M^*_{\la^{\prime}_c, \la^{\,}_d; \la^{\prime}_a, \la^{\prime}_b} \,
D_{\la^{\,}_C,\la^{\,}_C}^{\la^{\,}_c,\la^{\prime}_c}(z) \>, 
$$
and that the above equation leads to vanishing single spin asymmetries
\beq
A_N = \frac{ d\sigma^{\uparrow} - d\sigma^{\downarrow} }
           { d\sigma^{\uparrow} + d\sigma^{\downarrow} }
    = \frac{ d\sigma^{\uparrow} - d\sigma^{\downarrow} }
           { 2 \, d\sigma^{unp} } \>, \label{an}
\eeq
due to helicity conservation in the elementary interactions and  
parton collinear configurations in distribution and fragmentation 
functions (the notations used above should be self-explanatory, 
further details can be found, {\it e.g.}, in Ref. \cite{noi3}; $\uparrow$
and $\downarrow$ are directions perpendicular to the scattering plane).

Possible origins of single spin effects can be introduced by considering
$\bfk_\perp$ dependences in the quark distribution functions 
$\hat f_{q/p}(x, \bfk_\perp)$ \cite{siv}-\cite{boe} or in the quark 
fragmentation functions $\hat D_{h/q}(z, \bfk_\perp)$ \cite{col}; a 
consistent phenomenological approach has been developed taking these 
possibilities into account \cite{noi3}, \cite{noi1}, \cite{noi2} and 
assuming that the factorization theorem of
Eq.~(\ref{dsabpol}) holds also when intrinsic parton motion is included. 
At leading order in $k_\perp$ one has, for the $\pup p \to \pi X$ process:
\beq
\frac{E_\pi \, d^3\sigma^\uparrow} {d^{3} \bfp_\pi} 
- \frac{E_\pi \, d^3\sigma^\downarrow} {d^{3} \bfp_\pi} =
\sum_{a,b,c,d} \int \frac {dx_a \, dx_b} {\pi z} \> \times \label{gen} 
\eeq
$$
\Bigg\{ \int d^2 \bfk_{\perp} \,
\Delta^Nf_{a/\pup} (\bfk_{\perp}) \> f_{b/p} \,
\frac{d \hat \sigma} {d\hat t} (\bfk_{\perp}) \>
D_{\pi/c} \> +
$$
$$
\int d^2 \bfk'_{\perp}\, h_1^{a/p} \> f_{b/p} \>
\Delta_{NN} \hat\sigma(\bfk'_\perp) \>
\Delta^N D_{\pi/c}(\bfk'_\perp) \> +  
$$
$$
\int d^2 \bfk''_{\perp}\, h_1^{a/p} \> 
\Delta^Nf_{\bup/p} (\bfk''_{\perp}) \>
\Delta'_{NN} \hat\sigma(\bfk''_\perp) \>
D_{\pi/c}(z) \Bigg\} \>,
$$
where the first line in brackets corresponds to the so-called 
Sivers effect \cite{siv}, the second to Collins effect \cite{col} and the
third one to a recently proposed mechanism \cite{boe}. 

The new $\bfk_\perp$ and spin dependent functions appearing in Eq.~(\ref{gen})
have a partonic interpretation in terms of polarized quark distribution
and fragmentation functions as:
\beq
\Delta^N f_{q/\pup} = 
\hat f_{q/\pup}(x, \bfk_{\perp}) -  
\hat f_{q/\pdown}(x, \bfk_{\perp})\,, \label{delf}
\eeq
\beq
\Delta^N D_{h/q} = 
\hat D_{h/\qup}(z, \bfk_{\perp}) - 
\hat D_{h/\qdown}(z, \bfk_{\perp})\,, \label{deld}
\eeq
\beq
\Delta^N f_{\qup/p} = 
\hat f_{\qup/p}(x, \bfk_{\perp}) -  
\hat f_{\qdown/p}(x, \bfk_{\perp})\,. \label{delb}
\eeq

The other quantities appearing in Eq.~(\ref{gen}), apart from the unpolarized
quark distribution and fragmentation functions, $f$ and $D$, are the 
transverse spin content of the proton:
\beq
 h_1^{q/p} = f_{\qup/\pup}(x) - f_{\qdown/\pup}(x)
\eeq
and the elementary double spin asymmetries, computable in pQCD:
\beq 
\Delta_{NN} \hat\sigma = {d\hat \sigma^{\aup b \to \cu d} \over d\hat t} 
- {d\hat \sigma^{\aup b \to \cd d} \over d\hat t} \,, \label{dnn1}
\eeq
\beq 
\Delta'_{NN} \hat\sigma = {d\hat \sigma^{\aup \bup \to c d} \over d\hat t} 
- {d\hat \sigma^{\aup \bdown \to c d} \over d\hat t} \,\cdot \label{dnn2}
\eeq

Eq.~(\ref{gen}) can be used to obtain information on the new quantities
(\ref{delf})-(\ref{delb}) by fitting existing data on $\pup p \to \pi X$
\cite{e704}; once this has been done predictions for other processes can be 
given. This program has been partially performed in Refs. \cite{noi3},
\cite{noi1}, \cite{noi2} and \cite{noi4}. 

\section{Sivers and Collins effects alone}

In Ref. \cite{noi2} Sivers effect has been assumed to be the only origin
of single spin asymmetries and the corresponding expression of 
Eq.~(\ref{gen}) ({\it i.e.} its first two lines) has been used to fit
the data on $\pup p \to \pi X$, leading to an explicit expression for
$\Delta^Nf_{q/\pup}(x, \langle k_\perp \rangle)$ at an average intrinsic
transverse $k_\perp$ value; the data can be nicely fitted and the 
deduced expression of $\Delta^Nf_{q/\pup}(x, \langle k_\perp \rangle)$
leads to predictions for single spin asymmetries in $\bar p^\uparrow p
\to \pi X$ in agreement with data \cite{e704}. The resulting values of
$\Delta^Nf_{q/\pup}/2f_{q/p}$ also turn out to be reasonable and plausible.

A similar program, taking only Collins effect into account, has been
performed in Ref. \cite{noi3}, where the first and third line of 
Eq.~(\ref{gen}) have been used to explain the existing data and to
derive an explicit expression of $\Delta^ND_{\pi/q}(z, 
\langle k_\perp \rangle)$: again, data on $\pup p \to \pi X$ 
can be fitted, with some difficulties
at large $z$ values, and values of $A_N$ in $\bar p^\uparrow p \to \pi X$ 
processes are computed in agreement with experiment. In this case, however, 
the resulting values of $\Delta^ND_{\pi/q}/2D_{\pi/q}$ turn out to be 
at the border of acceptability, in that this ratio has to reach 1 at large
$z$ values in order to fit the data. 

It is clear that both Sivers and Collins effects might be at work at the 
same time and contribute in different proportions to the observed 
values of $A_N$. Not only: also the contribution suggested in 
Ref. \cite{boe} might be at work. Its consequences alone have not 
been studied yet, although it might be possible that the effects  
of $\Delta^Nf_{\qup/p}$ are evident in different $x_F$ regions from 
Sivers and Collins effects.

However, both the new, time-reversal odd functions $\Delta^Nf_{q/\pup}$
and $\Delta^Nf_{\qup/p}$ require some initial state interactions 
\cite{noi4}; these initial state interactions
are expected in hadron-hadron processes where many soft gluons
should be exchanged between the colliding particles. This is not so 
in lepton-hadron interactions where any extra exchange of photons 
gives only negligible corrections. Thus, we expect that only Collins 
effect might originate single spin asymmetries in lepto-production 
processes. We consider the most favourable situation, by
assuming that this is true also in $pp$ processes, and take the expression
of $\Delta^ND_{\pi/q}$ as determined in Ref. \cite{noi3}: under 
such an assumption -- only Collins effect at work to originate
single spin asymmetries in any inclusive process -- we are able to give 
predictions for $A_N$ in semi-inclusive lepton initiated processes.     
  
\section{Predictions for DIS processes}

We consider, in complete analogy to $\pup p \to \pi X$, the process
$\ell \pup \to \pi X$: the former process has an observed large $A_N$
and, if Collins mechanism is responsible for it, we also expect a large
asymmetry for the latter:
\beq
\frac{E_\pi \, d^3\sigma^\uparrow} {d^{3} \bfp_\pi} 
- \frac{E_\pi \, d^3\sigma^\downarrow} {d^{3} \bfp_\pi} =
\sum_q \int \frac {dx} {\pi z} 
\> \int d^2 \bfk_{\perp} \> \times \label{gendis} 
\eeq
$$
\left [ h_1^{q/p}(x) \> \Delta_{NN} \hat\sigma^q(x, \bfk_\perp) \>
\Delta^N D_{\pi/q}(z, \bfk_\perp) \right ]
$$
where
\beq 
\Delta_{NN} \hat\sigma^q = {d\hat \sigma^{\ell \qup \to \ell \qup} \over 
d\hat t} - {d\hat \sigma^{\ell \qup \to \ell \qdown} \over d\hat t} 
\> \cdot \label{dnn3}
\eeq

{}From the expressions of Ref. \cite{noi3} we can compute Eq.~(\ref{an})
and (\ref{gendis}) and we obtain indeed large values of $A_N$; 
in Fig.~1 we show results at $s$ = $52.5$ GeV$^2$
and $p_T = 1.2$ GeV/$c$ (typical HERMES values). Similar results hold 
at different energies \cite{noi4}.

\vspace{-14pt}

\begin{figure}[htb]
\begin{center}
\mbox{~\epsfig{file=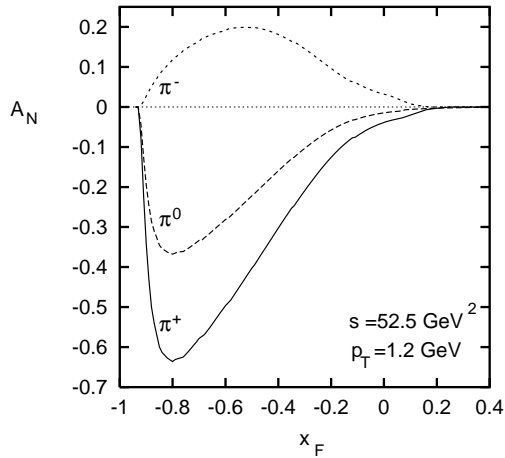,angle=-90,width=7cm}}
\vspace{-18pt}
\caption{$A_N(\ell p^{\uparrow}\to\pi X)$ for typical HERMES kinematics.}
\end{center}
\end{figure}

\vspace{-14pt}

A measurement of $A_N$ in $\ell \pup \to \pi X$ requires transversely 
polarized nucleons; however, single  
transverse spin asymmetries may be measurable also in the 
case of longitudinally polarized nucleons provided one looks at the double 
inclusive process, $\ell \pup \to \ell \pi X$ from which one can reconstruct 
the $\gamma^* \pup \to \pi X$ reaction, which, in general, occurs in 
a plane different from the $\ell-\ell'$ plane where the longitudinal 
nucleon spin lies; in this case one has (see Ref. \cite{noi4} for details):
\beq
\frac{d\sigma^{\gamma^* \pup \to \pi X}}{dx \, dQ^2 \, dz \, d^2p_T}
- \frac{d\sigma^{\gamma^* \pdown \to \pi X}}{dx \, dQ^2 \, dz \, d^2p_T}
= \sum_q \times \label{ancg}
\eeq
$$
h_1^{q/\pup} 
\left[ \frac{d\hat\sigma^{\gamma^* \qup \to \qup}} {dQ^2}
- \frac{d\hat\sigma^{\gamma^* \qup \to \qdown}} {dQ^2} \right] \>
\Delta^ND_{\pi/q}(p_T) \>.
$$
In Fig.~2 we show $A_N$ at the same energy values of Fig.~1, as
a function of $z$: it is large also in this case, although only at very 
large $z$ values which might be difficult to reach experimentally. 

\vspace{-14pt}

\begin{figure}[htb]
\begin{center}
\mbox{~\epsfig{file=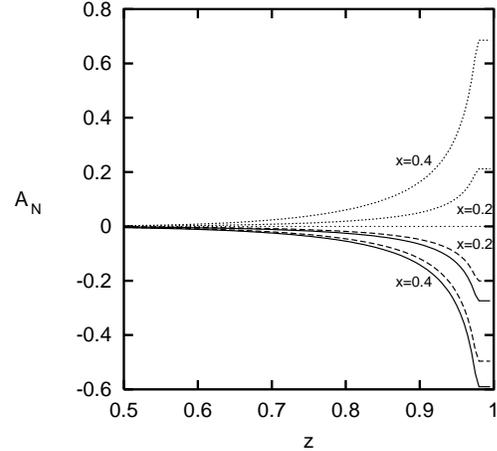,angle=-90,width=7cm}}
\vspace{-18pt}
\caption{$A_N(\gamma^*p^{\uparrow}\to\pi X)$: $s=52.5$ GeV$^2$, $Q^2 = 8$
GeV$^2$; solid, dashed and dotted line refer respectively to $\pi^+$, $\pi^0$
and $\pi^-$.}  
\end{center}
\end{figure}

\vspace{-14pt}

Single transverse spin asymmetries might be large in semi-inclusive 
DIS: their measurement is important and is feasible at several 
laboratories either with the existing configurations or with transversely
polarized future nucleon beams.

\end{document}